\documentclass
[10pt,aps,prc,twocolumn,showpacs,showkeys,amsmath,floatfix,superscriptaddress]
{revtex4}
\usepackage{color,graphicx}
\usepackage{mathptmx}                
\usepackage{dcolumn}                 
\usepackage{bm}                      
\newcommand{\cred}{\color{red}}

\begin{document}

\title{The heat capacity of the neutron star inner crust within an extended NSE model}  

\author{S. Burrello}
\affiliation{INFN, Laboratori Nazionali del Sud and Dip. di Fisica e Astronomia, Universit\`a di Catania, 95123 Catania, Italy}
\author{ F. Gulminelli}
\affiliation{CNRS and ENSICAEN, UMR6534, LPC, 14050 Caen c\'edex, France}
\author{ F. Aymard}
\affiliation{CNRS and ENSICAEN, UMR6534, LPC, 14050 Caen c\'edex, France}
\author{M.Colonna}
\affiliation{INFN, Laboratori Nazionali del Sud, 95123 Catania, Italy}
\author{Ad.R.Raduta}
\affiliation{NIPNE, Bucharest-Magurele, POB-MG6, Romania}

\begin{abstract}
\begin{description}
\item[Background] Superfluidity in the crust is a key ingredient for the cooling properties of proto-neutron stars. Present theoretical calculations employ the quasi-particle mean-field Hartree-Fock-Bogoliubov theory with temperature dependent occupation numbers for the quasi-particle states. 
\item[Purpose]  Finite temperature stellar matter is characterized by a whole distribution of different nuclear species. We want to assess the importance of this distribution on the calculation of heat capacity in the inner crust. 
\item[Method]  Following a recent work, the Wigner-Seitz cell is mapped into a model with cluster degrees of freedom. The finite temperature distribution is then given by a statistical collection of Wigner-Seitz cells. We additionally introduce pairing correlations in the local density BCS approximation both in the homogeneous unbound neutron component, and in the interface region between clusters and neutrons. 
\item[Results] The heat capacity is calculated in the different baryonic density conditions corresponding to the inner crust, and in a temperature range varying from 100 KeV to 2 MeV. We show that accounting for the cluster distribution has a small effect at intermediate densities, but it considerably affects the heat capacity both close to the outer crust and close to the core. We additionally show that it is very important to consider the temperature evolution of the proton fraction for a quantitatively reliable estimation of the heat capacity.
\item[Conclusions] We present the first modelization of stellar matter containing at the same time a statistical distribution of clusters at finite temperature, and pairing correlations in the unbound neutron component. The effect of the nuclear distribution on the superfluid properties can be easily added in future calculations of the neutron star cooling curves. A strong influence of resonance population on the heat capacity at high temperature is observed, which deserves to be further studied within more microscopic calculations.
\end{description}
\end{abstract}

\date{\today}


\maketitle

\section{Introduction}

Superfluidity in the crust is a key ingredient in the understanding of many different phenomena in compact star physics, from
the cooling of young neutron stars \cite{Lat94,Gne01}, 
to the afterburst relaxation in X-ray transients \cite{Pag12}, as well as in the understanding of glitches \cite{Pie14}.
Moreover, it is well-known that pairing correlations reduce the crust thermalization time by a large fraction \cite{Gne01,For10}.
The specificity of the inner crust is the simultaneous presence of clusters and homogeneous matter, which are both influenced
by pairing interactions. Indeed the occurrence of dishomogeneities has a non-negligible influence on the pairing properties of the inner crust
\cite{For10,Bar98,San04,Cha10,Pas15}, and consequently on the time evolution of the surface temperature of the neutron star.

Present studies of crust superfluidity at finite temperature are typically done solving Hartree-Fock-Bogoliubov (HFB) equations
in the so-called Wigner-Seitz approximation \cite{Neg73}, meaning that the assumption is done that the cluster component is
given by a single representative quasi-particle configuration, corresponding to a single representative nucleus immersed in a neutron 
gas.
These works do not consider the fact that at finite temperature a wide distribution of nuclei is
expected to be populated at a given crust pressure and temperature conditions. 
Moreover, at the extremely low proton fractions associated to the inner crust, deformed nuclear structures and beyond drip-line light nuclear resonances can participate to the statistical equilibrium, and might be too exotic to be well described through standard mean-field calculations. 
Non-spherical pasta structures, which are not accessible to the spherical mean-field, 
 have been reported to be only marginally populated in $\beta$-equilibrium \cite{Ava08}. However, at sufficiently high temperatures, light particles 
can appear and even become dominant in the composition of matter \cite{Roe13,Ava12} and can modify the local distribution of 
neutron density, and the associated pairing field.

A way to include these beyond-mean field effects is given by  finite temperature Nuclear Statistical
Equilibrium (NSE) models. In the most recent NSE implementations  
\cite{Hec09,Hem10,Rad10,Rad14,Fur13,Buy13} 
the distribution of clusters  is taken into account and obtained self-consistently under conditions of statistical
equilibrium. In some of these models both the gas-cluster interaction and the self-interaction of the gas are included, though within semi-classical approximations
\cite{Gul15}. In particular in Ref.\cite{Gul15}  it is shown that a proper definition of the cluster self-energies in the NSE cluster distribution
allows recovering the zero temperature limit of a single Wigner-Seitz approximation
in the (Extended) Thomas-Fermi limit. 

This kind of approaches are however not adequate to describe the heat capacity of the crust because they 
 do not consider the presence of pairing correlations.
The aim of this paper is to analyze how the non-homogeneity of crust matter and the associated wide distribution of nuclear
species, affects the superfluid properties of the crust. Specifically, we introduce pairing correlations both in the cluster
and homogeneous matter component of the NSE model in the local BCS approximation and study the effect of the cluster 
distribution on the heat capacity of the inner crust. We will show that the single nucleus approximation is perfectly adequate in some regions
of the inner crust, but non-negligible effects of the cluster distribution are seen close to the drip point, and close to the crust-core transition.

In most HFB calculations for the cooling problem \cite{For10,Bar98,San04,Cha10,Pas15}, the approximation is made  that the proton fraction does not evolve with the temperature
and can be estimated by the value imposed, at each baryonic density, by the condition of neutrino-less chemical equilibrium at 
zero temperature of reference calculations \cite{Neg73}. 
Even with the inclusion of pairing, the NSE model is still much less numerically demanding than a full HFB calculation at finite
temperature. For this reason, we have released this approximation and imposed  $\beta$-equilibrium at each finite temperature. This condition is justified by the fact that the time scale of cooling is sufficiently slow to insure the chemical equilibrium of weak processes at all times \cite{Gne01}.
The temperature dependence of the proton fraction is shown to have considerable effects 
on the heat capacity.

The paper is organized as follows.
The improved NSE model with inclusion of pairing correlations in the neutron gas,  is presented in section \ref{sec:NSE}.
In this section are detailed the superfluid neutron gas 
 (section \ref{sec:gas}) and cluster distribution (section \ref{sec:clus}) modelling, 
the calculation  of the total energy (section \ref{sec:ene}) and the calculation of the in-medium modification of the cluster surface and pairing properties due to
the presence of the gas  (section \ref{sec:inmed}).
Section \ref{sec:results} is devoted to the presentation of the results. 
The composition of the inner crust in terms of cluster distribution and unbound neutrons, 
 the temperature evolution of the energy and the heat capacity are given in the dedicated 
subsections (III A and III B). Section \ref{sec:mass} discusses the importance of a highly predictive model for the binding energies of the different nuclear species. 
Finally section \ref{sec:concl} gives a summary and conclusions.

\section{The improved Nuclear Statistical Equilibrum model} \label{sec:NSE}

The complete formalism that we use can be found in Refs.\cite{Gul15,Rad14}. 
Here, we recall the main equations and detail the 
inclusion of pairing both in the bulk and in the surface region inside the Wigner-Seitz cells, which was not considered in Ref.\cite{Gul15}.

The model is based on a statistical distribution of 
compressible nuclear 
clusters immersed in a homogeneous background of self-interacting nucleons and electrons.
We label each nuclear species composed by $N$ neutrons and $Z$ protons by their mass number and 
bulk asymmetry $(A,\delta)$.
Even below drip, the asymmetry in the bulk for a nucleus in the vacuum, 
$\delta_0$ differs from the global asymmetry of the nucleus, $I=1-2Z/A$, because of the 
presence of a neutron skin and  Coulomb effects.
The relation between $\delta_0$ and  $I$ 
is given by~\cite{Mye80,Cen98,Cen09}:
\begin{equation}
\delta_0 = \frac{I+\frac{3 a_C}{8 Q} \frac{Z^2}{A^{5/3}}} {1+\frac{9 E_{sym}}{4 Q} \frac{1}{A^{1/3}}},
\label{paperpana1:eq:deltacl}
\end{equation}
where $E_{sym}
$ is the symmetry energy at saturation,  
$Q$ is the surface stiffness coefficient extracted from a semi-infinite nuclear matter calculation
and $a_C$ is the Coulomb
parameter taken equal to $a_C=0.69$ MeV.
The continuum states leading to the existence of a free nucleon gas can in first approximation be modelled 
as leading to a constant density contribution. As a consequence,
the bulk asymmetry inside the clusters can be decomposed into the asymmetry of the gas
 $\delta_g$ weighted by the gas fraction $x_{gc}=\rho_{g}/\rho_0$  
inside the cluster, plus the asymmetry of the cluster in the vacuum
$\delta_{0}$ weighted by the complementary mass fraction 
$x_{cl}=(\rho_0-\rho_{g})/\rho_0$, namely 
\begin{equation}
\delta = \left( 1-\frac{\rho_{g}}{\rho_0}\right) \delta_{0}
+\frac{\rho_{g}}{\rho_0} \delta_{g} 
\label{eq:deltain}
.\end{equation}
In the previous equation $\rho_0$ denotes the bulk density.
Variational arguments lead to the conclusion that, independent of the presence of an external gas, the
equilibrium bulk density corresponds to the saturation density at the corresponding bulk asymmetry\cite{Pan13}.
This means that the following expression, at the second order in asymmetry, can be used for  the bulk density:
\begin{equation}
\rho_{0} = \rho_{sat} \left( 1-\frac{3 L_{sym} \delta^2}{K_{sat}+K_{sym} \delta^2 }  \right),
\label{eq:rho0}
\end{equation}
In this equation, $\rho_{sat}$ is the saturation density of symmetric nuclear matter, 
$L_{sym}$ and $K_{sym}$ are the slope and curvature of the symmetry energy at saturation.
Then, solving the coupled Eqs.(2) and (3), it is possible to extract the bulk density and asymmetry.
\subsection{The free energy of the superfluid gas} \label{sec:gas}

\begin{figure}[htbp]
\includegraphics[width=0.9\linewidth]{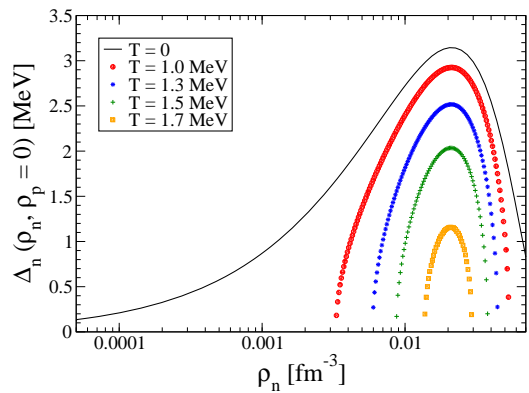}
\caption{
(Color online) $^1S_0$ pairing gap as a function of density for homogeneous neutron matter at zero temperature as obtained 
from Brueckner-Hartree Fock calculations (full line from ~\cite{Cao06}). The figure also shows the energy gap deduced
solving the BCS gap equations at finite temperature (symbols from~\cite{Bur14}).
}
\label{fig_gap}
\end{figure}

The energy density of a nuclear gas, of density $\rho_g$ and asymmetry $\delta_g = 1 - 2\rho_{gp}/\rho_g$, at finite temperature T, in the mean field approximation reads \cite{Chamel,Bur14}
(q=n,p):
%
\begin{equation}
\epsilon_{HM}(\rho_g,\delta_g) = g_q\sum_q \int_0^\infty \frac{dp}{2\pi^2 \hbar^3} p^2 f_q \frac{p^2}{2m^*_q}+\mathcal{E}_{pot},
\label{eq:HM}
\end{equation}
with
\begin{equation}
\mathcal{E}_{pot} = \mathcal{E}_{sky} + 
\frac{1}{4}\sum_{q=n,p} v_\pi (\rho_{gq})\tilde{\rho}^*_{gq} \tilde{\rho}_{gq} .
\end{equation}
Hereafter the acronym HM stands for Homogeneous Matter.
We will use the Sly4 parametrization \cite{Cha98} 
of the Skyrme energy functional for the local energy density $\mathcal{E}_{sky}$ 
and the effective nucleon mass $m^*_q$,
 for the numerical applications of this paper.
In Eq.(\ref{eq:HM}),
$g_q=2$ is the spin degeneracy in spin-saturated matter 
and $f_q$ is the 
particle occupation number:
\begin{equation}
f_q=\frac 1 2  \left [ 1 - \frac{\xi}{E_\Delta} tanh \left ( \frac{E_\Delta}{2 T}\right ) \right ],
\end{equation}
with $E_\Delta = \sqrt{\xi^2+\Delta^2}$ and $\xi=\epsilon_q - \mu_q = p^2/2m^*_q - \tilde{\mu}_q$.
$\mu_q$ and $\tilde{\mu}_q$ denote, respectively, chemical potential and reduced chemical
potential, and 
\begin{equation}
\epsilon_q=\frac{p^2}{2m^*_q(\rho_g,\delta_g)} +\frac{\partial \mathcal{E}_{pot}}
{\partial \rho_{gq}} (\rho_g,\delta_g) 
\label{new}
\end{equation}
is the single particle energy. 
$\Delta$ is the temperature dependent  pairing gap and
$\tilde{\rho}_{gq}=2\Delta(\rho_{gq})/v_\pi(\rho_{gq})$ denotes the anomalous 
density.  In Eq.\ref{new}, the derivative with respect to  $\rho_{gq}$
is taken at constant $\tilde{\rho}_{gq}$.

The pairing interaction is given by \cite{Chamel,Bur14} :
\begin{equation}
v_{\pi q}(\rho_{gq})=V_\pi \left [ 1-\eta \left ( \frac{2\rho_{gq}}{\rho_{sat}} \right ) ^\alpha \right ],
\end{equation}
where the parameters $V_\pi$, $\eta$, $\alpha$
are fixed imposing to reproduce the $^1S_0$ pairing gap of pure neutron matter as obtained in Brueckner-Hartree-Fock calculations \cite{Cao06}. The resulting gap is displayed in Fig.\ref{fig_gap}.
Then the density dependence of the pairing strength can be calculated exactly in the BCS approximation by inverting
the gap equation
\begin{equation}
1=-v_\pi(\rho_{gq}) \int_0^{p_\Lambda} \frac{dp}{2\pi^2\hbar^3} p^2 \frac{1}{2\xi} \left [ 1- 2f_q(p) \right ].
\end{equation}
The reduced chemical potential  $\tilde{\mu}_q$ is moreover obtained  by fixing the particle
number density:
\begin{equation}
\rho_{qg}= g_q \int_0^\infty \frac{dp}{2\pi^2\hbar^3} p^2 f_q.
\end{equation}

It should be noticed
that, owing to the zero range of the pairing interaction,
a cutoff has to be introduced in the gap equations to
avoid divergences. Following Refs.\cite{Chamel,Bur14}, 
we adopt the energy cutoff $p^2_\Lambda/2m^*_q -  \tilde{\mu}_q = 16$ MeV.

The free energy density is obtained adding the entropy term in the mean-field approximation:

\begin{eqnarray}
f_{HM}(\rho_g,\delta_g)&=& \epsilon_{HM}(\rho_g,\delta_g)
- T s_{HM}(\rho_g,\delta_g),
\label{eq:f_HM}
\end{eqnarray}
where the entropy density is given by :
%
\begin{eqnarray}
s_{HM}(\rho_g,\delta_g)&=&-\sum_q g_q\int_0^\infty \frac{dp}{2\pi^2\hbar^3} p^2    [ n_q \ln n_q + \nonumber \\
&+& \left ( 1 - n_q \right ) \ln \left ( 1 - n_q \right ) ]
\label{eq:s_HM}
\end{eqnarray}
with $n_q(\epsilon_q)=(1+\exp (E_\Delta(\epsilon_q)/T))^{-1}$.

\subsection{The cluster distribution} \label{sec:clus}

A given thermodynamic condition in terms of temperature, 
baryonic density and proton fraction $(T,\rho_B,y_p)$ is characterized by a mixture of configurations defined by $k=\{V_{WS}^{(k)},A^{(k)},\delta^{(k)},\rho_g,\delta_g\}$ with a free energy given by \cite{Gul15}:
\begin{eqnarray}
F_{WS}^{(k)}&=&F_\beta(A^{(k)},\delta^{(k)},\rho_g,\delta_g) + V^{(k)}_{WS} f_{HM} (\rho_g,\delta_g) \nonumber \\
& +&  V^{(k)}_{WS} f_{el} (\rho_p) ,
\label{fenergy}
\end{eqnarray}
In this expression, $f_{el}$ is the electron free-energy density,
$\rho_p$ is the total proton density and $V^{(k)}_{WS}$ denotes the Wigner Seitz volume.  $F_\beta$ is the free energy of the cluster immersed in the nucleon gas:
\begin{eqnarray}
F_\beta(A,\delta,\rho_g,\delta_g)&=&
E^{vac}(A,\delta) - T \ln  \left (A_e^{\frac 3 2} c_\beta V_t \right ) \nonumber \\
&+&\delta F_{bulk} \nonumber \\
&+&\delta F_{surf}+\delta F_{Coul}
\label{fenergy_cl_ws}
\end{eqnarray}
where the total volume $V_t$ has been introduced and we have defined the bound fraction of the cluster by 
$A_e=A\left (1-\rho_g/\rho_0\right )$, $Z_e=Z\left (1-\rho_{gp}/\rho_{0p}\right )$,
with $\rho_{0p}=\rho_0(1-\delta)/2$. 

The temperature dependent degeneracy factor includes the sum over the cluster excited states as:
\begin{equation}
c_\beta=\left (\frac{mT}{2\pi\hbar^2} \right)^{3/2} \int _0^{<S>}dE \left [ \rho_{A,\delta}(E)
\exp(-E/T) \right ],  \label{deg_bucurescu}
\end{equation}
where $\rho_{A,\delta}$ is the density of states of the cluster,
$<S>=\min(<S_n>,<S_p>)$ is the average particle separation energy, $m$ is the nucleon mass. See Ref.\cite{Gul15} for details. 

We can observe that the cluster energy is modified with respect to the corresponding vacuum energy
$E^{vac}$  both because of nuclear and Coulomb in-medium effects. 

The modification of the nuclear free energy consists of a bulk term 
\begin{equation}
\delta F_{bulk}=- f _{HM}(\rho_g,\delta_g) 
V_{cl}
, \label{eq:dfbulk}
\end{equation}
due to the presence of the gas in the same spatial volume, $V_{cl} = \frac{A}{\rho_0(\delta)}$, occupied by the cluster, 
and a
surface term $\delta F_{surf}$ which accounts for the isospin-dependent modification of the surface tension 
due to the presence of the gas at the surface of the cluster. The calculation of this last term will be detailed in section \ref{sec:inmed}. Moreover, to insure additivity of the cluster and the gas component, only the bound 
part of the cluster $A_e$ appears in the translational entropy term of Eq.(\ref{fenergy_cl_ws}), which also can be considered as an in-medium effect.


The screening effect of the electron density $\rho_e=\rho_p$ which neutralizes 
the Wigner-Seitz cell leads to a modification
of the cluster free energy according to:
\begin{equation}
\delta F_{Coul}= a_c  f_{WS}(\rho_p,\rho_{0p})A^{5/3}\frac{(1-I)^2}{4},
\label{eq-coul}
\end{equation}
with the Coulomb screening function in the Wigner-Seitz approximation:
%
\begin{equation}
  f_{WS}(\rho_p,\rho_{0p})=\frac 32\left ( \frac{\rho_p}{\rho_{0p}}\right )^{1/3} -
\frac 12\left ( \frac{2\rho_p}{\rho_{0p}}\right ). \label{eq:screening}
\end{equation}
%

The volume of the Wigner-Seitz cell associated to each nuclear species is univocally defined by the charge conservation constraint:
%
\begin{equation}
\rho_p=\rho_e=\frac{Z_e}{V_{WS}} + \rho_{gp}
\end{equation}
leading to
\begin{equation}
V_{WS}=\frac{Z}{\rho_{0p}} \frac{\rho_{0p}-\rho_{gp}}{\rho_{p}-\rho_{gp}} \label{V_WS}.
\end{equation}

The equilibrium distribution is obtained by minimizing 
 the total free energy corresponding to an arbitrary collection of different cells $k$,
subject to the constraint of total baryonic and charge density conservation \cite{Gul15}:
\begin{eqnarray}
\rho_B&=&\frac{\sum_k  n^{(k)}(A_e^{(k)}+V_{WS}^{(k)}\rho_g )}
{\sum_k  n^{(k)} V_{WS}^{(k)}}, \label{eq1} \\
\rho_p&=&\frac{\sum_k  n^{(k)}(Z_e^{(k)}+V_{WS}^{(k)}\rho_{gp} )} {\sum_k  n^{(k)} V_{WS}^{(k)}}.
\label{eq2}
\end{eqnarray}

The result is a NSE-like expression for the cluster multiplicities \cite{Gul15}:

\begin{eqnarray}
\ln n^{(k)}  
&=&
-\frac{1}{T}\left (
F_\beta(A^{(k)},\delta^{(k)},\rho_g,\delta_g) 
-\mu_B A_e - \mu_p  Z_e
\right ), \label{mult_nse}
\end{eqnarray}
where the chemical potentials can be expressed 
as a function of the gas densities only:

\begin{eqnarray}
\mu_B &\equiv& \frac{\partial f_{HM}}{\partial \rho_g};  \label{chem1}\\
\mu_p &\equiv& \frac{\partial f_{HM}}{\partial \rho_{gp}} \label{chem2}.
\end{eqnarray}

The numerical solution of Eqs.(\ref{eq1}),(\ref{eq2}) for the two unknown $\rho_g$, 
$\rho_{gp}$ closes the model.

Concerning the cluster binding energies 
$E^{vac}(A,\delta)$, 
theoretical coherence with the treatment of the gas demands that they are evaluated with the same 
Skyrme energy functional employed for the gas component. We will use for the binding energy the analytical expressions proposed in Ref.\cite{Dan09}:
\begin{equation}
E^{vac}=a_v A -a_s A^{2/3}-a_a(A) A I^2-a_c  A^{5/3}\frac{(1-I)^2}{4}, \label{dan}
\end{equation}
with the asymmetry energy coefficient:
\begin{equation}
 a_a(A) =\frac{a_v^a}{1+\frac{a_v^a}{a_s^a A^{1/3}}},
\end{equation}
where the different parameters are fitted from numerical Skyrme calculations in slab geometry \cite{Dan09}. 
The pairing contribution to the cluster energy is evaluated according to the phenomenological
expression: $E_{pair} = \pm \Delta_{pair}(A) = \pm 12/\sqrt{A}$,
where the +(-) sign refers to even-even (odd-odd) nuclei.

 It is important to observe that this formula,Eq.(\ref{dan}), similar to any other mean-field model,
systematically underbinds light particles, which will then tend to be underestimated in the calculations.
We will discuss the effect of this limitation in section \ref{sec:mass}.
Concerning the density of states $\rho_{A,\delta}(E)$, mean-field models are known to be far off in the reproduction of these observables, and empirical adjustements have to be done. For this reason we use a back-shifted Fermi gas model with parameters fitted from experimental data \cite{Buc05}.

\subsection{Computation of the total energy} \label{sec:ene}
 
 For the calculation of the heat capacity, the simplest approximation consists in considering the contribution
of the cluster and the gas as simply additive, which corresponds to neglecting the term $\delta F_{surf}$ in Eq.(\ref{fenergy_cl_ws}). 

In some early studies \cite{Lat94, Piz02}, 
the cluster contribution was completely ignored, and the nonuniform distribution
was replaced with a uniform gas formed by the total number of neutrons in the cell or by taking only the
number of the unbound neutrons. It was shown in Ref.\cite{For10} that these approximations very poorely reproduce the total heat capacity of a complete HFB calculations: the shape of the peak is too sharp, and the transition temperature is underestimated. 

The displacement of the transition temperature can be simply understood as an effect of the gas density, 
which for a given particle number and cell size is obviously modified in the presence of the cluster, because this latter occupies a finite volume. This effect cannot be simply accounted if the same boundary conditions of the complete HFB calculations are applied to the uniform neutron 
gas configuration \cite{For10}. 

To compute the contribution of the gas to the energy density in the simplified hypothesis 
$\delta F_{surf}=0$, we have to consider the total volume, $V_g$, accessible to the gas, i.e. the volume left
after excluding the volume of the clusters, and evaluate the corresponding gas
volume fraction $x_g=\lim_{V_t \to \infty} \frac {V_{g}}{V_t}$.
Then the gas energy density can be simply written as: $\epsilon_g = 
x_g~\epsilon_{HM}$. 
In the Single Nucleus Approximation (SNA) employed in HFB calculations, we can consider a single representative Wigner-Seitz cell and write
$x_g = 1-V_{cl}/V_{WS}$.

In our model, the full distribution of clusters is accounted for 
and the volume fraction $x_g$ accessible to the gas results:

\begin{equation}
x_g=\lim_{V_t \to \infty} \frac {
V_{g}}{V_t} = 1 -\lim_{V_t \to \infty}\frac{1}{{V_t}}\sum_{k} n^{(k)} \frac{A^{(k)}}{\rho_0(\delta^{(k)}) }, \label{eq:vexcl}
\end{equation}
where the total volume $V_t$ can be written as $V_t = n_{tot}\langle V_{WS}\rangle$, being  $\langle V_{WS}\rangle$  the average size of the Wigner-Seitz volume and $n_{tot} = \sum_{k}n^{(k)} $ the total cluster multiplicity.  Indicating with 
$p^{(k)} = n^{(k)}/n_{tot}$ the normalized cluster multiplicity, 
%
%
the contribution of the gas to the energy density is therefore:
\begin{equation}
\epsilon_{g}=\epsilon_{HM}\left ( 1- 
\frac{1}{\langle V_{WS}\rangle } \sum_{k} p^{(k)}  \frac{A^{(k)}}{\rho_0(\delta^{(k)})} \right )  
\end{equation}
and the total baryonic energy density of star matter is
\begin{eqnarray}
\epsilon_{tot}=\epsilon_{g}+\epsilon_{cl}&=&
\epsilon_{g} + \frac{1}{\langle V_{WS}\rangle}\sum_{k} p^{(k)}\langle E(A^{(k)},\delta^{(k)})\rangle \label{eq:en_nosurface}
\end{eqnarray}

The energy $E(A,\delta)$ 
entering Eq.(\ref{eq:en_nosurface}) is given by the vacuum energy, Eq.(\ref{dan}),
shifted by the electron screening effect, $\delta E_{Coul}$, and
augmented of the average translational energy, $3/2~T$, and average excitation energy $\langle E^* \rangle$ 
corresponding to the considered temperature and cluster 
density of states \cite{Gul15}: 
\begin{eqnarray}
\langle E (A,\delta,\rho_p,T)\rangle&=&E^{vac}(A,\delta)  +\delta E_{Coul}(A,\delta,\rho_p) \nonumber \\
&+&\frac 32 T + \langle E^*(A,\delta,T) \rangle.  \label{eq:e_t}
\end{eqnarray}
Here, $\delta E_{Coul}=\delta F_{Coul}$ (Eq.(\ref{eq-coul})) because 
the Coulomb shift is determined by the electrons. These
latter being independent fermions with respect to the nucleons,
they have no effect on the baryonic state counting.
 
As shown by Eq.(\ref{fenergy_cl_ws}),
this simple excluded volume effect can be also formulated as the additivity of the gas with the 
bound part of the clusters. 

This decomposition of the Wigner-Seitz cell between cluster and gas, 
accounting for the excluded volume effect, was tried in the recent HFB analysis of Ref.\cite{Pas15}.
It was shown that the transition temperature of the superfluid gas is correctly recovered, but the peak is
still sharper than in the full HFB calculation. Moreover, in the full calculation a second peak at higher temperature 
can appear in the outer part of the inner crust, depending on the energy functional \cite{For10,Pas15}.
This peak corresponds to the temperature at which the whole system becomes non-superfluid, due to the pairing 
effect in the surface of the cluster. It is clear that this peak cannot be reproduced in the hypothesis of energy additivity.

These observations show the importance of accounting for the in-medium pairing corrections of the interface between the cluster and the gas, that we examine in the next section.

\subsection{In medium effects} \label{sec:inmed}

In Eq.(\ref{eq:en_nosurface}) we have assumed that the bound part of the cluster and the gas contribution are additive.
This hypothesis is based on the approximation that : (i) the most  important part of the in-medium 
correction is given by the Coulomb screening 
by the electron gas, and by the Pauli-blocking effect of high energy cluster 
single particle states due to the gas ~\cite{Typ10}; (ii) this latter effect can be 
approximately accounted for by subtracting from the local energy density the 
contribution of the unbound gas states. The Coulomb screening is indeed considered in the functional $E(A,\delta)$,
and the bulk part of the in medium correction is accounted for by considering the cluster excluded volume,
Eq.(\ref{eq:vexcl}). 
 
The approximation of Eq.(\ref{eq:en_nosurface}) neglects the modification of the cluster surface tension due to the presence 
of an external neutron gas.

This residual  in-medium modification of the cluster energy $\delta E_S$ can be computed by subtracting 
to the total energy in each Wigner-Seitz cell the contribution of the gas alone and of the nucleus alone, following~\cite{Aym14}:
\begin{eqnarray}
\delta E_{s}&=&E^{tot}-E(A,\delta,\rho_p,T) + \nonumber \\
&-& \left ( V_{WS}-\frac{A}{\rho_0(\delta)}\right ) 
\epsilon_{HM} (\rho_g,\delta_g),
\label{total_energy_modification}
\end{eqnarray}
Considering that this correction is expected to be a surface effect, it appears reasonable to compute it in the local density 
approximation (LDA), $\delta E_{s}\approx \delta E_{LDA}$.
 %
%
%

Since the proton contribution to the nucleon gas is very small for beta-equilibrated matter and in the temperature regime concerned by our study,
we can safely neglect any Coulomb effect to  $\delta E_{LDA}$, meaning that we can consider solely the nuclear part of the energy in Eq.(\ref{total_energy_modification}).
Then the cluster energy $E$ in the LDA can be decomposed in an isospin dependent  bulk part and residual terms varying with $A$ slower than linear (surface, curvature and higher order): 
\begin{equation}
E_{LDA}(A,\delta)=\frac{\epsilon_{HM} (\rho_{0},\delta)}{\rho_{0}} A  +E_S .
\label{energy_nucleus_vacuum}
\end{equation}
A similar decomposition can be applied to the total LDA energy :
\begin{eqnarray}
E_{LDA}^{tot}&=&\int_ 0^{R_{cl}}   \epsilon_{HM} (\rho(r),\delta(r)) d^3r  \nonumber \\
&&\hspace{1cm}+ \int_{R_{cl}}^{R_{WS}}  \epsilon_{HM} (\rho(r),\delta(r)) d^3r  \nonumber  \\
&=&  \epsilon_{HM} (\rho_{0},\delta) V_{cl} + 
 \epsilon_{HM} (\rho_g,\delta_g)\left (V_{WS}-V_{cl} \right ) \nonumber \\
 &&\hspace{1cm} + E_{S,m}.  
\label{emedium}
\end{eqnarray}
where $R_{WS}$ is the radius of the Wigner-Seitz cell, 
$R_{cl}$ is the hard-sphere radius, associated with the cluster volume $V_{cl}$, 
and $E_{S,m}$ represents 
a surface term since the bulk parts have been highlighted. 

Using Eqs.~(\ref{total_energy_modification}), (\ref{energy_nucleus_vacuum}) and (\ref{emedium}),
we can express the  residual in-medium modification simply as  
\begin{equation}
\delta E_{s}(A,\delta,\rho_g,\delta_g,T) =  E_{S,m}-E_S. \label{eq:desurf}
\end{equation}
Since $\delta E_S$ is 
related to the two surface terms deduced from Eqs.~(\ref{energy_nucleus_vacuum}) and (\ref{emedium}),
we can expect
the following relation to hold: $\delta E_S = c_{s}A^{2/3}$,
where the temperature dependent parameter $c_s$ should have a weak dependence on $A$,  revealing the small effect of the curvature terms.

These in-medium corrections were evaluated in Ref.\cite{Aym14} adding to the LDA also higher orders in $\hbar$ in the semiclassical Thomas-Fermi developement of the energy functional, but neglecting the pairing interaction and the temperature dependence. 
It was shown that $\delta E_S$ is indeed a surface term $\propto A^{2/3}$, but it  displays a very complex behavior with the cluster bulk asymmetry $\delta$, the gas density $\rho_g$ and the gas asymmetry $\delta_g$.

 In this work we include the temperature effect and the pairing interaction according to Eq.(\ref{eq:HM}), but we limit ourselves to the simple LDA. Gradient and spin-orbit terms are therefore neglected in the surface correction.  

In order to evaluate the in-medium surface correction through Eq.(\ref{emedium}), a model for the density profiles $\rho(r)$, $\delta(r)$ has to be assumed. 
We use the simple Wood-Saxon analytical profiles proposed in Ref.\cite{Pan13} and successfully compared to full Hartree-Fock calculations in spherical symmetry in Refs.\cite{Pan13,Aym14}:
 \begin{eqnarray}
\rho(r) &\equiv& \frac{\rho_{0}-\rho_{g}}{ 1+\exp (r-R)/a}+\rho_{g} \\
\rho_{p}(r) &\equiv& \frac{\rho_{0p}-\rho_{gp}}{ 1+\exp (r-R_p)/a_p}+\rho_{gp} 
\label{eq:FD-gas}
\end{eqnarray}
such that the local asymmetry is given by $\delta(r)=1-\rho_p(r)/\rho(r)$.
The radius parameters $R,R_p$ entering the density profile~(\ref{eq:FD-gas}) are related to the equivalent  
hard sphere radii by
\begin{equation} 
R = R_{cl} \left[ 1 - \frac{\pi^2}{3} \left(\frac{a}{R_{cl}}\right)^2 \right],
\label{paperpana1:eq:radiusws}
\end{equation} 
and a similar relation holds for $R_p$.
The diffuseness parameters $a, a_p$ of the  total density profile are assumed to depend quadratically on the bulk asymmetry $\delta$, 
$a_i=\alpha_i+\beta_i\delta^2$, where $\alpha_i$ and $\beta_i$ were fitted from HF calculations in Ref.\cite{Pan13}.
 
Using Eq.(\ref{eq:FD-gas}) the in-medium surface correction can be finally expressed as
\begin{eqnarray}
\delta E_{s} &= & \int_ 0^{R_{WS}} d^3r
\left [  \epsilon_{HM} (\rho(r),\delta(r)) - \epsilon_{HM} (\rho_{cl}(r),\delta_{cl}(r))  \right ] \nonumber \\
& -& \epsilon_{HM}(\rho_g,\delta_g) \left ( V_{WS} - \frac {A} {\rho_0(\delta)} \right ),
\label{deltaE}
\end{eqnarray}
where $\rho_{cl}(r)$ and $\rho_{p,cl}(r)$ are the total and proton densities that correspond to the same $(A,\delta)$ 
cluster in the absence of the gas
 \begin{eqnarray}
\rho_{cl}(r) &\equiv& \frac{\rho_{0}}{ 1+\exp (r-R)/a} \\
\rho_{p,cl}(r) &\equiv& \frac{\rho_{0p}}{ 1+\exp (r-R_p)/a_p} ,
\label{eq:FD-clus}
\end{eqnarray}
and  $\delta_{cl}(r)=1-2\rho_{p,cl}(r)/\rho_{cl}(r)$.

For the low temperatures which are of interest in the present study,
the in-medium surface energy  correction computed here is 
expected to give a small effect  to the composition of the inner crust \cite{Rad14}. 
The effect of the in-medium correction will therefore be estimated perturbatively.  
We  assume that, for a given thermodynamic condition $(\rho_B,y_p, T)$, the in-medium surface correction 
$\delta E_S(A,\delta,\rho_g,\delta_g)$ affects only slightly the gas density and composition, and consequently 
the chemical potentials. This correction will then be taken using the values for $\rho_g,\delta_g$ obtained from a NSE calculation where the in-medium effect 
is not considered. With this assumption,  the modified binding energies solely depend on the cluster and on the thermodynamic condition
and can therefore be simply added a-posteriori to the energy density. 

The final expression for the total baryonic energy density at finite temperature is then given by:
\begin{eqnarray}
\epsilon_{tot}&=&
\epsilon_{g} + \frac{1}{<V_{WS}>}\sum_{k} p^{(k)}[\langle E(A^{(k)},\delta^{(k)})\rangle \nonumber \\
&+&\delta E_S(A^{(k)},\delta^{(k)})], \label{eq:en_surface}
\end{eqnarray}
where all terms depend on the temperature, and on the gas density and composition.

\begin{table*}[tb]
\caption{From left to right are given: the total baryonic density, the proton fraction at T=100 keV, the proton fraction considered in Ref.\cite{For10}, the gas density at T=100 keV, the gas density obtained at T=0 in Ref.\cite{For10}, the radius of the average Wigner-Seitz volume calculated at T=100 keV and the radius of the cell at T=0 shown in Ref.\cite{For10}.}

{\renewcommand\arraystretch{1.2}
\begin{ruledtabular}
\begin{tabular}{c|ccccccc}
Cell & 
$\rho_B$ [fm$^{-3}$] & $y_p^0$ & 
$y_p^{\textup{HFB}}$ & $\rho_g^0$ [fm$^{-3}$]& ${\rho}_g^{\textup{HFB}}$ [fm$^{-3}$] & $\langle R_{WS}^0 \rangle $ [fm] & $R_{WS}^{\textup{HFB}}$  [fm]\\ 
\hline
$1$   &  $4.8 \times 10^{-2}$  &  $0.032$  &  $0.027$  &  $3.9 \times 10^{-2}$  &  $3.8 \times 10^{-2}$  &  9   &  20 \\
$2$   &  $2.0 \times 10^{-2}$  &  $0.035$  &  $0.028$  &  $1.7 \times 10^{-2}$  &  $1.7 \times 10^{-2}$  & 22  &  28 \\
$3$   &  $9.0 \times 10^{-3}$  &  $0.040$  &  $0.037$  &  $7.5 \times 10^{-3}$  &  $7.5 \times 10^{-3}$   & 30  &  33 \\
$4$   &  $5.8 \times 10^{-3}$  &  $0.045$  &  $0.045$  &  $4.8 \times 10^{-3}$  &  $4.6 \times 10^{-3}$  & 33  &  36 \\
$5$   &  $3.7 \times 10^{-3}$  &  $0.054$  &  $0.053$  &  $3.0 \times 10^{-3}$  &  $3.0 \times 10^{-3}$  & 36  &  39 \\
$6$   &  $1.6 \times 10^{-3}$  &  $0.083$  &  $0.080$  &  $1.2 \times 10^{-3}$  &  $1.1\times 10^{-3}$   & 41  & 42 \\
$7$   &  $9.0 \times 10^{-4}$  &  $0.122$  &  $0.125$  &  $5.4 \times 10^{-4}$  &  $5.3 \times 10^{-4}$  & 44  & 44 \\
$8$   &  $6.0 \times 10^{-4}$  &  $0.162$  &  $0.160$  &  $2.8 \times 10^{-4}$  &  $2.8 \times 10^{-4}$  & 46  & 46 \\
$9$   &  $4.0 \times 10^{-4}$  &  $0.220$  &  $0.200$  &  $1.2 \times 10^{-4}$  &  $1.3 \times 10^{-4}$  & 47  & 49 \\
$10$ &  $2.8 \times 10^{-4}$  &  $0.284$  &  $0.222$  &  $2.8 \times 10^{-5}$  &  $7.4 \times 10^{-5}$  & 48  & 54 \\ 
\end{tabular}
\end{ruledtabular}}
\label{table:cells}
\end{table*}

\section{Results} \label{sec:results}


In order to facilitate a quantitative comparison with the previous literature, we have chosen 
ten representative values for the baryonic density which have been proposed in the seminal
paper by Negele and Vautherin \cite{Neg73}. These values cover the inner crust of the neutron star,
approximately from the emergence of the neutron gas close to the drip point (cell 10) to a density 
close to the crust-core transition (cell 1), where bubbles and possibly other exotic nuclear shapes start
to be formed. We recall that such structures are not included in our model. 
The corresponding values of the baryonic density, as well as the gas density, the proton fraction and the radius of the average Wigner-Seitz cell volume we obtain
imposing the $\beta$-equilibrium condition, at the lowest temperature (T=100 KeV) considered in this study, are given in Table \ref{table:cells}. 
We notice that the proton fraction increases, whereas the gas density decreases moving from
cell 1 to cell 10. 
For comparison, proton fraction, gas density and radius of Wigner-Seitz cell obtained at T=0 in the full HFB calculation of Ref.\cite{For10} are also given in the table.
In Ref.\cite{For10}, the same Sly4 parametrization 
was used in the calculations.  
As far as the gas density is concerned, the difference between
the HFB values and our results, at the lowest temperature considered, are of the order of $2\%$ or less, except
for the lowest densities; in that case however the gas contribution is negligible. 
It should also be noticed that the 
HFB calculations of  Ref.\cite{For10} adopt the same proton fraction of the representative calculations of
\cite{Neg73}, i.e. the $\beta$-equilibrium condition is not consistently implemented. 
The residual variation can be partly due to the different energetic description of the clusters. Our simplified mass model from Ref.\cite{Dan09} is augmented of a phenomenological pairing term \cite{Gul15} but 
does not contain shell effects. Neutron shell effects do not play any role above drip, but proton shell closures
are known to be still effective at zero temperature in the inner crust \cite{Pea12}, which can slightly affect the neutron gas density close to the drip condition. 
More important, the pairing interaction of this work is not the same as in Ref.\cite{For10}. As explained in section 
\ref{sec:gas}, we have fitted the parameters of the pairing interaction from ab-initio BHF calculations of infinite 
neutron matter at zero temperature. 
This choice, also employed in Refs.\cite{Bur14,Pas15}, is justified by the fact that the dominant pairing contribution comes from the unbound neutrons, which constitute, at the thermodynamic limit of the neutron star, an homogeneous neutron matter system. The cluster-gas interface, which is treated in the present work in the local density BCS approximation (see section \ref{sec:inmed}) , gives only a correction to this dominant term. Conversely, in the finite Wigner-Seitz calculation of Ref.\cite{For10}, these parameters were fitted from finite nuclear properties \cite{San04}. The resulting maximum pairing gap $\Delta_{max}\approx 3$ MeV is very close to the one displayed in Fig.\ref{fig_gap}, but the density dependence of the pairing gap (see Ref.\cite{For10}) is different with respect to our calculation. 
Concerning the radius of the average volume of the Wigner-Seitz cell, again our results are in good agreement with HFB, except at the highest density (Cell1). As it will be shown further in the paper, this difference is due to the dominance of light resonances in our calculation, which are not included in a mean-field approach.

\subsection{Composition  of the inner crust} \label{sec:thermo}
Most thermodynamic calculations of the inner crust \cite{For10,Bar98,San04,Cha10,Pas15} 
neglect the temperature variation of the proton 
fraction due to the temperature dependence of the chemical potentials entering the neutrinoless $\beta$-equilibrium condition:
\begin{equation}
\label{eq:beta_eq}
\mu_n(T)-\mu_p(T)=\mu_e(T)
.\end{equation}
\begin{figure}[htbp]
\includegraphics[width=0.8\linewidth]{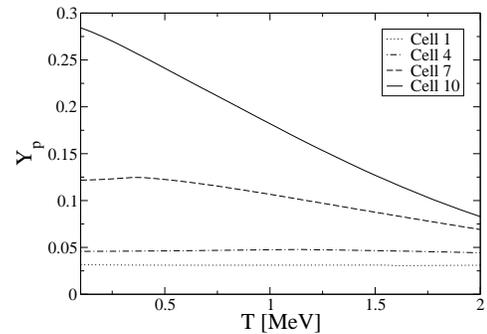}
\caption{
Temperature evolution of the global proton fraction obtained by imposing the neutrinoless 
$\beta$-equilibrium condition \eqref{eq:beta_eq} for four representative cells.  
}
\label{fig:yp}
\end{figure}

\begin{figure}[htbp]
\includegraphics[width=0.8\linewidth]{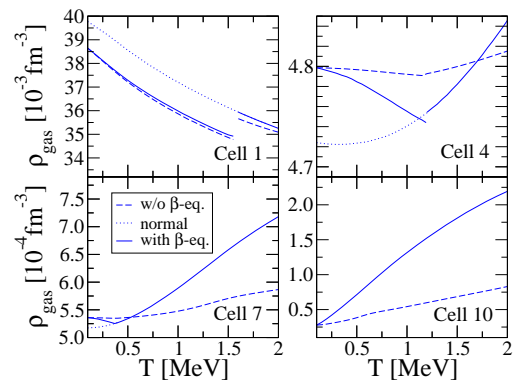}
\caption{
(Color online) Temperature evolution of the unbound gas component in the same representative cells as in Fig. \ref{fig:yp}. Full line: complete NSE calculation. Dashed line: the value of the global proton fraction is assumed equal to the one calculated from $\beta$-equilibrium at the lowest temperature, $y_p(T)=y_p(0.1 $MeV$)$. Dotted line: as the full line, but neglecting the pairing 
interaction.}
\label{fig:rhogas}
\end{figure}

This variation, as obtained in our calculations,  is shown in Fig.\ref{fig:yp},
in four representative cells spanning the density and temperature interval concerned by this study.
 We can see that the change of the proton fraction is indeed very small close to the crust-core transition (up to cell 4),
but it cannot be neglected at lower densities (cells 5 to 10). 
The density corresponding to the unbound neutron component is shown in Fig.\ref{fig:rhogas} 
for the same baryonic density conditions as in Fig.\ref{fig:yp}.

The impact of the $\beta$-equilibrium condition on the gas density can be appreciated from the difference between the full and the dashed curve in Fig.\ref{fig:rhogas}. 
It is clear from this result that the $\beta$-equilibrium condition has to be consistently implemented at each temperature.
However, the most striking feature of Fig.\ref{fig:rhogas} is the clear discontinuity 
observed at the highest densities (up to cell 4 in the present calculation), corresponding to the transition point
from superfluid to normal matter.  
 
At first sight it is surprising to observe a density discontinuity, which is characteristic of first order phase transitions,
at the superfluid-normal fluid transition, which is second order. 
This behavior is due to the fact that we are not observing an equation of state, that is $\rho(T)$ at constant chemical potential, but a specific thermodynamic transformation implied by the minimization of the system total
free energy.
Specifically, one should consider that the pairing gap jumps, continously but suddenly, to zero at the critical temperature. This behavior
influences the energetics of the system and may create discontinuities in the solution obtained for the gas density. 
This is particularly evident in the cells where, as in Cell 1, the gas density is larger than the
value associated with the maximum gap (see Fig.1). In this case, 
the gas density solution corresponding to zero temperature in the full NSE calculation is lower than the gas density 
obtained neglecting the pairing interaction (dotted line in 
Fig.\ref{fig:rhogas}), because it corresponds to a larger 
gap energy. 
As the temperature increases, in the regime where pairing is still active, the gas density decreases because the (negative)
pairing contribution to the gas energy reduces.   
At the critical temperature the non-superfluid solution is recovered as it should. This corresponds to a higher density value, leading to a discontinuity.

\begin{figure}[htbp]
\includegraphics[width=0.9\linewidth]{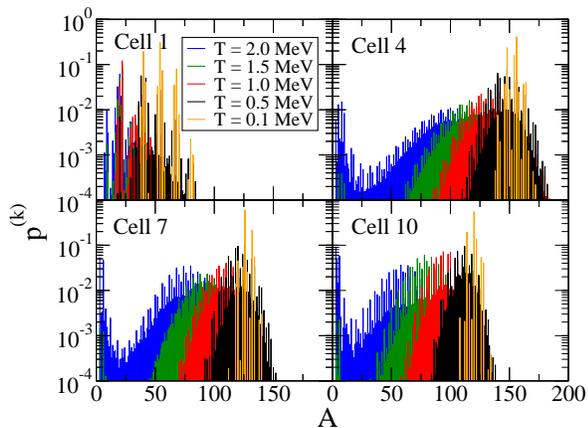}
\caption{
(Color online) Normalized cluster size distribution at different temperatures in the same four representative cells as in Fig. \ref{fig:yp} and at $\beta$-equilibrium.
}
\label{fig:yields}
\end{figure}

The distribution of the cluster size as a function of the temperature 
is displayed  in Fig.\ref{fig:yields}.

We can see that at the lowest densities and temperatures the distribution is strongly peaked and can be safely approximated by a unique nucleus, but increasing the temperature and/or moving towards the inner part of the crust, many different nuclear species can appear with comparable probability. 
Moreover, light particles systematically dominate at the highest temperatures. Such configuration cannot be addressed in mean-field based formalisms like HFB.
%
\begin{figure}[htbp]
\includegraphics[width=0.9\linewidth]{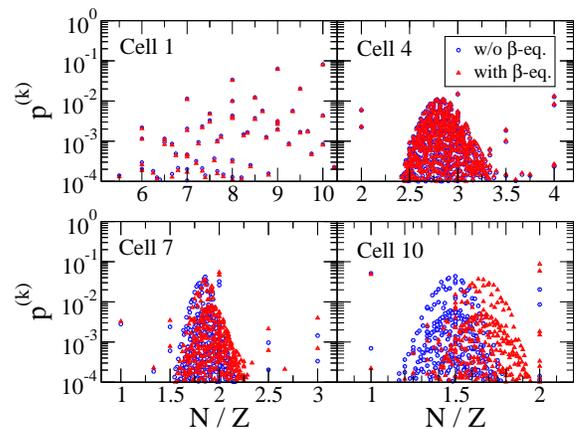}
\caption{Normalized isotopic cluster distribution in the same four representative cells as in Fig.\ref{fig:yp}, as obtained at the highest 
temperature considered, $T = 2$ MeV. The different symbols indicate 
results at $\beta$-equilibrium (triangle) or assuming a global proton fraction equal to that one corresponding to $\beta$-equilibrium but at the lowest temperature, $y_p(T)=y_p(0.1 $MeV$)$.}
\label{fig:NZ}
\end{figure}
Fig.\ref{fig:NZ} shows the cluster isotopic distribution, for the same four cells, at the temperature T=2 MeV.  
Results obtained neglecting the proton fraction variation imposed by the $\beta$-equilibrium condition are also shown.  
One can observe that, especially for the lowest density cells (Cells 7 and 10), the cluster asymmetry is significantly larger
when the $\beta$-equilibrium condition is imposed.
It is also interesting to notice that, at the high limits of the N/Z distribution, 
the yield 
is higher than the corresponding value obtained in absence of $\beta$-equilibrium.
These extreme N/Z values are obtained from the lightest clusters,
which dominate at the temperature considered (see Fig.\ref{fig:yields}).
Properly accounting for the $\beta$-equilibrium thus increases the contribution of the most unbound clusters.

From these results we can already anticipate that neglecting the temperature evolution of $\beta$-equilibrium will lead to a strong underestimation of the energy density, and the associated heat capacity, at high temperature.

\subsection{ Energy and heat capacity} \label{sec:heat}

\begin{figure}[htbp]
\includegraphics[width=0.9\linewidth]{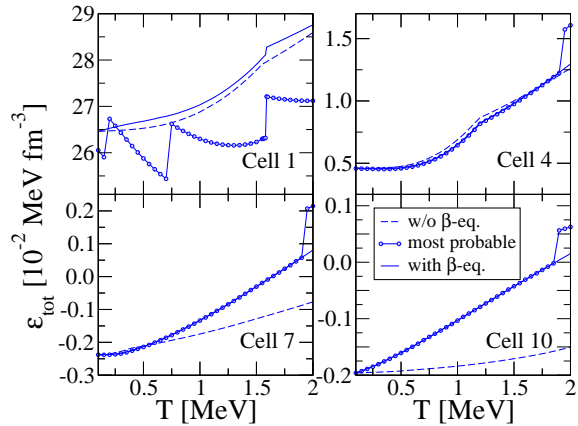}
\caption{
(Color online) Temperature evolution of the baryonic energy density for the same representative cells as in Fig.\ref{fig:yp}. Full line: complete NSE calculation. Dashed line: as the full line, but the value of the global proton fraction is assumed equat to that one calculated from $\beta$-equilibrium  at the lowest temperature, $y_p(T)=y_p(0.1 $MeV$)$. Lines with symbols: as the full line, but the NSE distribution is replaced with the most probable Wigner-Seitz cell.
}
\label{fig:ene}
\end{figure}

The variation with temperature of the energy density is displayed, for the same density conditions as in the previous figures, in Fig. \ref{fig:ene}.

The effect of the temperature dependence of the $\beta$-equilibrium condition can be appreciated comparing the full thin lines with the dashed lines. 
As expected, we can see that the temperature evolution of the proton fraction has a strong effect on
the energy density, especially at the lowest densities.  

Finally, the lines with symbols give the energy density of the most probable Wigner-Seitz cell, to be compared to the complete result (full lines) where the whole distribution of cells is taken into account.
We can see that the effect of properly accounting for the cluster distribution is very important at the highest densities, but also 
at the lowest ones when the temperature gets higher. Indeed these situations are
dominated by the emergence of light clusters. Close to the crust-core transition, the matter is so neutron rich that standard heavy clusters are not favored any more with respect to more exotic neutron-rich forms of matter. As it can be seen from Fig.\ref{fig:yields}, in this thermodynamic conditions the mass distribution extends up to $A\approx 100$ 
but is dominated by light resonances at the limit of the nuclear binding (heavy hydrogen, helium, or lithium).
The energy density associated with the full distribution is thus very different from the one associated with the most probable cluster. Specifically, the discontinuities observed in the most probable Wigner-Seitz cell
in Fig.\ref{fig:ene} (upper left) appear at the temperatures where a transition occurs between the different elements. 

It is important to remark that in this density region in principle non-spherical pasta phases, which are not included in the present work,  could dominate over the light resonances. This is certainly true 
for low temperatures and matter close to isospin symmetry
since the breaking of spherical symmetry leads to an important gain in binding energy \cite{Wat00}. 
However finite temperature calculations in $\beta$-equilibrium \cite{Ava12} tend to show that non-spherical pasta phases are only marginal, meaning that the 
energy behavior displayed in 
Fig.\ref{fig:ene} might be physical.

A similar transition, from heavy cluster dominated to light resonance dominated configurations, is observed at all densities.
We recall that starting from cell 2 the density is too low for pasta phases to be present. This transition, leading to a sharp discontinuity in the energy density of the most probable Wigner-Seitz cell, physically corresponds to the melting of clusters inside a hot medium. In a mean-field treatment, cluster disappearence can only lead to a homogeneous medium, because small wavelength fluctuations cannot be treated in these approaches. However such fluctuations are entropically favored and naturally appear in the NSE treatment at high temperature.

The transition temperature from the superfluid to the normal fluid phase is signalled by a kink in the behavior of the energy density, which will lead to a peak in the associated heat capacity. This transition
 occurs at the same point in the full NSE calculation and considering only the most probable Wigner-Seitz cell.
This can be understood from the fact that the electron and nucleon gases are uniform along the different cells, meaning that by construction the density and isospin caracteristics of the gas are the same in the two calculations.  
On the other hand, it is interesting to observe that a temperature shift could be observed if a standard calculation considering a single representative cell (SNA)  was performed \cite{Gul15},
as in the well-known Lattimer-Swesty model \cite{Lat91}.
Indeed the baryonic density associated to the Wigner-Seitz cell of the most probable cluster is not the same as the total baryonic density of the distribution. 
This is a consequence of the fact that, especially at high temperature,  the most probable cluster can be very different from the average cluster, 
thus it is very important to consider the full cluster distribution, as in the NSE calculations.

This also means that the consideration of the cluster distribution could modify the transition temperature as predicted by finite temperature HFB, though the effect is expected to be small.  

\begin{figure}[htbp]
\includegraphics[width=0.9\linewidth]{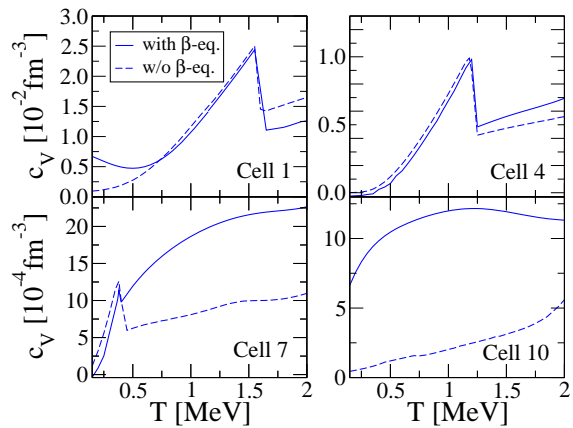}
\caption{
(Color online) Temperature evolution of the heat capacity for the same representative cells as in Fig.\ref{fig:yp}.  Full line: complete NSE calculation. Dashed line: as the full line, but the value of the global proton fraction is assumed equal to that one calculated from $\beta$-equilibrium  at the lowest temperature, $y_p(T)=y_p(0.1 $MeV$)$.  
}
\label{fig:cv}
\end{figure}

Fig.\ref{fig:cv}  shows the temperature behavior of the total baryonic energy  derivative with respect to temperature, in four different cells.  
The temperature derivative was performed numerically following the trajectory of $\beta$-equilibrium: this means that only the total baryonic density is constant, but the proton fraction is not. 
As we have anticipated observing the energy density behavior of Fig.\ref{fig:ene}, the temperature dependence of the $\beta$-equilibrium condition is seen to have a dramatic effect on the heat capacity.
In particular the peak due to the phase transition is strongly smeared out in the outer region of the inner crust, from cell 7 to 10, due to the rapid variation of the unbound component with temperature implied by the $\beta$-equilibrium condition (see Fig.\ref{fig:rhogas}).
On the contrary, at the highest densities (cells 1 to 3) the consideration of the temperature variation of the proton fraction increases the size of the peak. Indeed, in this case the $\beta$-equilibrium path favors a discontinuous trend of all thermodynamic quantities at the transition point (see Fig.\ref{fig:rhogas}).

The LDA approximation was compared to HFB calculations in the case of trapped fermionic atoms in~\cite{Gra03}. It was shown that this approximation nicely works even in small systems $A\approx 50$ at zero temperature, but it rapidly deteriorates  at finite temperature. This is expected from the Ginzburg-Landau theory in cases where the critical temperature is much higher than the harmonic level spacing. In particular the LDA pairing field is seen to show a sudden drop at the surface, which is not apparent in the full HFB.

We however expect this limitation of LDA to be less severe in our case,  because contrary to Ref.~\cite{Gra03} we do not use the LDA to solve the variational problem, but only to calculate the energy correction. Moreover our physical system is obviously not the same as in~\cite{Gra03}. We have verified that in our Wigner-Seitz cells the radial profile of the pairing field does not drop off but presents a decreasing tail, similar to HFB results.

\begin{figure}[htbp]
\includegraphics[width=0.8\linewidth]{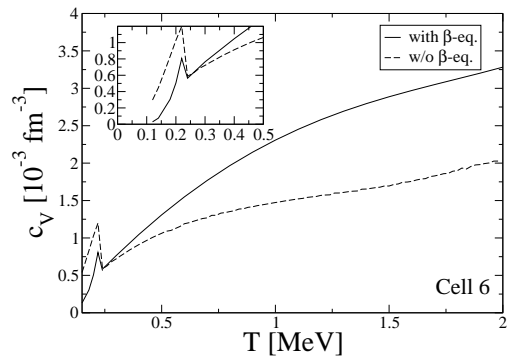}
\caption{
Temperature evolution of heat capacity for Cell 6, obtained
considering a pairing interaction with the same parameters of Ref.\cite{For10}. Full line: complete NSE calculation. Dashed line: as the full line, but the value of the global proton fraction is assumed equal to that one given in Ref.\cite{For10} (see also Table I). 
The inset shows a zoom at low temperature in order to facilitate the comparison with the results of Ref.\cite{For10}.
}
\label{fig:cvfortin}
\end{figure}

Concerning the heat capacity, as shown in Fig.\ref{fig:cv}, its quantitative value cannot be directly compared to the results of 
previous HFB works \cite{For10,Bar98,San04,Cha10,Pas15} because of the different mean-field and/or pairing model, and because of the non-negligible effect of the cluster distribution that we have observed in Fig.\ref{fig:ene}. However, we have verified that the temperature location of the heat capacity peak, its height and width are almost identical to the results of Ref.\cite{For10}, if we take the same parameters for the pairing interaction employed in that work.
This is illustrated in Fig.\ref{fig:cvfortin}, where we represent the corresponding results for the heat capacity, obtained 
imposing the $\beta$-equilibrium condition or neglecting it (as in the HFB calculations). 
It is observed that the full curve compares rather well with the results of Ref.\cite{For10}.
It is also interesting to notice that, as already pointed out,  the consideration of  $\beta$-equilibrium induces
non negligible effects on the $c_V$.  

\subsection{The effect of mass functionals} \label{sec:mass}

In all the calculations presented in the previous sections, 
we have systematically used the Skyrme-based liquid-drop formula, Eq.(\ref{dan}).
This choice allows a consistent treatment of the bound and unbound matter component within the same energy functional. However,  light clusters are systematically underbound with respect to heavier ones.
To give an example, employing the parameters extracted in \cite{Dan09} for Sly4, the binding energy of a $\alpha$ particles is underestimated of $\Delta B/B=20~ \%$ while it is overestimated of  $\Delta B/B=13~\%$ for $^{208}Pb$.
This effect is even more dramatic for the most neutron rich light resonances, at the limit of nuclear binding, which can in principle be excited in the extremely neutron rich $\beta$-equilibrated matter of proto-neutron stars at finite temperature: the last bound hydrogen isotope is $^3H$ according to 
the simplistic formula Eq.(\ref{dan}), while controlled extrapolations from experimental mass measurements predict that $^7H$ should be bound by $6.58$ MeV \cite{Aud12}. 

In Figs. \ref{fig:yields} and \ref{fig:ene} we have seen that at sufficiently high temperature, the last bound isotopes of light elements can become dominant in the composition of matter. It is therefore interesting to see how much these results depend on the poor energy description of light clusters
of our mass formula.
We have therefore repeated the same calculations, replacing Eq.(\ref{dan}) with the experimental value of the binding energy, whenever this value is known \cite{Aud12}. 
By the very definition of the inner crust, all the nuclei populated with non-negligible probability in the different density and temperature conditions explored in this work are beyond the dripline. 
This means that their experimental binding energy is typically not known, and Eq.(\ref{dan}) is still used
for those nuclei in the new calculation.
However experimental or extrapolated mass values exist for all bound isotopes of the lightest elements
$Z\leq 3$, and in that case the experimental value is used. 

\begin{figure}[htbp]
\includegraphics[width=0.8\linewidth]{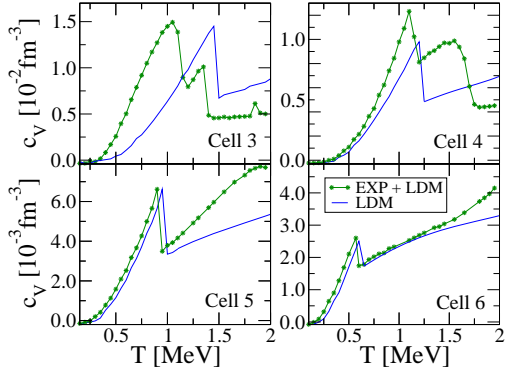}
\caption{
(Color online) Temperature evolution of heat capacity for four intermediate cells. Full lines: complete NSE calculation making use of the analytical expressions, Eq.(\ref{dan}), for binding energies (labeled as LDM in the figure). Line with symbols: as the full line, but experimental binding energies, from \cite{Aud12}, are used whenever available.
}
\label{fig:cv_exp}
\end{figure}
%

We find that, in the range of temperatures considered,  the results are similar to the ones presented in Fig. \ref{fig:cv} both for the highest (cells 1-2) and lowest (cells 7 to 10) densities. 
This means that the underbinding of light clusters does not influence the heat capacity calculation.

\begin{figure}[htbp]
\includegraphics[width=0.9\linewidth]{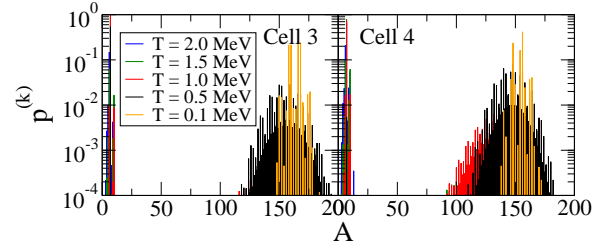}
\caption{
(Color online) Normalized cluster size distribution at different temperatures, as obtained from complete NSE calculation using experimental binding energies, from \cite{Aud12}, whenever available. Results are shown for two representative cells where the transition to a dominance of light clusters is observed in the
range of temperatures considered. 
}
\label{fig:na_exp}
\end{figure}

However, as shown in Fig.\ref{fig:cv_exp},
in the cells from 3 to 6 the situation is very different and the effect of accounting for the 
experimental binding energy of light clusters has a dramatic 
consequence. 
Indeed we see that accounting for the whole distribution of Wigner-Seitz cells, including the contribution of light clusters and resonances, modifies the height of the  heat capacity peak and also its position in temperature.
Moreover, an extra peak appears, which was not present in the calculations of 
Fig. \ref{fig:cv}. 
This peak corresponds to the ``critical'' temperature of the light clusters, depending on the thermodynamical conditions of the cell. 
This effect is easy to understand: the temperature at which the nuclei melt into a gas of free particles and resonances depends on the energy of these latter. If resonances correspond to bound states, they will dominate over the standard nuclei component at much lower temperatures than if they lie high in the continuum. 
The dominance of light clusters and resonances
induces a change in the temperature dependence of the energy density, leading to an additional peak in the $c_V$.
To better illustrate this point, Fig.\ref{fig:na_exp} shows the cluster distribution obtained considering the experimental binding energies, whenever available,  
in the case of cells 3 and 4, where the second peak in the heat capacity is observed.
One can appreciate that the cluster distribution is quite different with respect
to the results shown in Fig.\ref{fig:yields}.  Moreover, we observe that the location of the second peak of the heat capacity, shown in Fig.\ref{fig:cv_exp}, 
coincides with the temperature where the cluster distribution starts to be dominated by light clusters. 
It should also be noticed that the same features could also appear in cells at
lower densities, but at temperature values that are beyond the range considered
in the present study.  

However, a few words of caution, about employing the experimental masses, are in order. 
All the calculations presented in this chapter have been obtained including in the statistical weight of the clusters the bulk part of the in-medium free energy shift (Eq.(\ref{eq:dfbulk})), while the surface contribution 
(Eq.(\ref{eq:desurf})) has been added a-posteriori perturbatively, in order to consider the smearing effect of the density distribution on the pairing field. 
Contrary to bulk in-medium effects which increase the binding energy of the cluster, surface interaction with the surrounding gas is strongly dependent on the cluster asymmetry, as well as on the density and proton fraction of the gas.
Surface in-medium shifts for very neutron rich species immersed in a neutron gas tend in particular to decrease the binding energy of the cluster~\cite{Rad14}; it is therefore possible that a self-consistent inclusion of this energy term in the statistical calculation will reduce the contribution of the light resonances.
Moreover, our local density BCS approximation to evaluate the pairing contribution of an inhomogeneous density distribution, including the population of light resonances, is certainly a quite crude approximation.

\begin{figure}[htbp]
\includegraphics[width=0.9\linewidth]{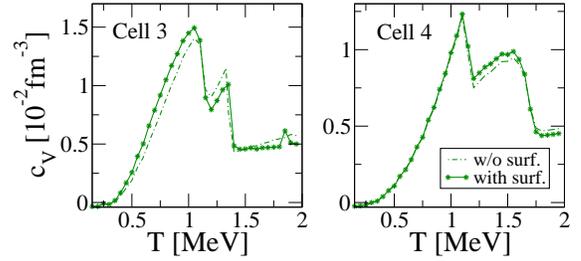}
\caption{
(Color online) Temperature evolution of heat capacity for the same two cells as in Fig.\ref{fig:na_exp}. Full line with symbols: complete NSE calculation using experimental binding energies, from \cite{Aud12}, whenever available. Dashed line: as the full line but neglecting in-medium effects.
}
\label{fig:cv_surf}
\end{figure}
The effects of the surface corrections, as described in Section \ref{sec:inmed}, are 
evidenced in Fig.\ref{fig:cv_surf}. 
The comparison between the full and dashed lines allows appreciating the importance of the density fluctuations inside the Wigner-Seitz cells. 
In the calculation illustrated by dashed lines, the total energy is simply given by the sum of the cluster and uniform gas component according to Eq.(\ref{eq:en_nosurface}), as suggested in early papers \cite{Lat94, Piz02}. 
The energy contribution of the cluster-gas interface, according to Eq.(\ref{eq:en_surface}), is considered in the full NSE results, 
shown by full lines. 

It should be noticed that the effect of density fluctuations is never negligible, but still represents a correction of the total energy density, 
thus globally justifying the perturbative treatment developed in section \ref{sec:inmed}.
As mentioned above, further corrections could be necessary in the case 
of very neutron rich species immersed in a neutron gas \cite{Rad14}. This point
is currently under study. 

In the present calculations we observe a small, though appreciable, effect
of the in-medium corrections on the heat capacity. 
In particular, we notice that, especially in the calculations neglecting the
surface effects (dashed line),  the transition temperature from superfluid to normal matter is very sharply defined.
This is due to the fact that the gas density is characterized by a single value at each temperature point. This artificial feature comes from the neglect of the neutron density distribution inside the Wigner-Seitz cells. Our procedure to
introduce surface corrections can partially 
cure this problem and leads to the results represented by the full lines. We can see that the transition temperature is smeared, as expected, and as it is observed in HFB calculations \cite{For10}.
Moreover a third small peak appears, in cell 3, at $T\approx$ 1.8 MeV, 
due to the disappearance of pairing effects on the surface of the clusters.

\section{Conclusions} \label{sec:concl}

In this paper we have presented a calculation of the heat capacity in the inner crust of proto-neutron star, within an approach based on cluster degrees of freedom that considers the complete distribution of different nuclear species in thermal and $\beta$-equilibrium. Superfluidity is taken into account including the pairing contribution of the homogeneous unbound neutron component in the BCS approximation.  A standard pairing interaction is employed, with parameters fitted such as to reproduce the pairing gap of infinite neutron matter, as calculated from ab-initio Brueckner-Hartree-Fock calculation. A non-relativistic Skyrme energy functional is used for the mean-field part of the unbound nucleon energy, as well as for the energy functional of the clusters.  In this modelization, interparticle interactions are explicitly accounted for the unbound particles, and implicitly for the bound ones through the cluster energy functional. 
Interactions between bound and unboud particles are taken into account by
considering the 
modification of the cluster surface due to the presence of the nucleon
gas, which in turn affects the pairing properties. 
To this purpose, we have introduced an interface correction calculated in the local density BCS approximation. 
The resulting heat capacity appears compatible with complete HFB calculations as far as the location and the width of the peak associated with the transition from superfluid to normal matter is concerned, if the same pairing interaction is employed and the same treatment of the $\beta$-equilibrium condition is performed. 
Indeed we show that an accurate treatment of $\beta$-equilibrium is important for a quantitative determination of the heat capacity, and consequently the neutron star cooling curve. Specifically, accounting for the temperature dependence of proton fraction is seen to modify the energy density in a sizeable way, and to sharpen the phase transition peak close to the crust-core interface.

The added value of the present semiclassical modelling with respect to more sophisticated HFB calculation in the Wigner-Seitz cell is the consideration of the full distribution of different nuclear species at finite temperature with their proper statistical weight.
We show that this feature considerably affects the heat capacity. In particular, the cluster disappearence at high temperature does not lead in this model to a uniform gas of nucleons, but correlations are still present in the form of exotic neutron-rich resonant states at the limit of nuclear binding. 
This feature may lead to the appearance of an extra peak in the heat capacity, corresponding to the cluster ``critical'' temperature, at which the nuclei melt into a gas of free particles and resonances. 
Moreover, 
treating the pairing properties of this inhomogeneous matter in the local density approximation leads to the prediction of a second peak in the heat capacity associated to the superfluid-normal fluid transition of this clusterized matter. A similar feature was already reported in the literature \cite{For10,Pas15} with HFB: a second peak is observed when the critical temperature is attained for the cluster surface. 
Further calculations within a more microscopic treatment are needed to confirm this finding.

\section*{Acknowledgments}
This work has been partially funded by New-Compstar, COST Action MP1304.

\end{document}